Georgia Southern University
College of Information Technology

Reduction of Field Loss by a Video Processing System

**Dr. Timur Mirzoev**
Department of Information Technology
College of Information Technology
P.O. Box 8150
Statesboro, GA 30460

Introduction

Streaming of 60 de-interlaced fields per second digital uncompressed video with 720x480 resolution without a loss of video fields is one of the desired technologies by scientists in biomechanics. If it is possible to stream digital uncompressed video without dropped video fields, then a sophisticated computer analysis of the transmitted via IEEE 1394a connection video is possible. Such process is used in biomechanics when it is important to analyze athletes' performance via streaming digital uncompressed video to a computer and then analyzing it using specific software such as Arial Performance Analysis Systems. Unfortunately, when 60 de-interlaced fields per second digital uncompressed video is streamed to a computer, some video fields maybe lost and not stored on a laptop computer successfully. If the streamed video does not contain all video fields, then it is not possible to utilize it for further analysis and; important decisions resulting from the desired video may not occur. This study emphasizes and relates to biomechanics as the loss of video fields becomes an essential limitation in digital video processing when biomechanical computer analysis is performed. The purpose of this research was to identify an optimal computer hardware model that results in reducing the number of factors that contribute to the loss of video fields when 60 de-interlaced fields per second digital uncompressed video streaming is performed over IEEE 1394a connection.

Motion analysis in biomechanics provides important tools for scientists to evaluate and analyze movements that create opportunities for improvements in sports performance. High quality video streaming is already widely available with conventionally wired camcorder to a computer. However, when DV-quality video is being transmitted to a laptop computer, video fields are lost which makes the video digitizing process impossible to be completed correctly.

In order to determine which hardware components or how many video streaming sources contribute to the loss of video fields during a DV video transmission over IEEE 1394a connection, there are several important considerations that needed to be analyzed. First, appropriate hardware and software setup is essential for providing DV-quality video streaming for biomechanical applications. Second, it is essential to study and examine hardware factors and video streaming sources that contribute to loss of video fields which, if the loss occurs, limits a scientist's ability to digitize and analyze captured video for biomechanical analysis. Along with the hardware design, the software package Ariel Performance Analysis System (APAS) motion analysis software was used as a tool for capturing video and performing biomechanical analyses on a computer.

Besides providing benefits to biomechanical specialists, a study which examines the hardware components that may contribute to the loss of video fields would also benefit a university's undergraduate and graduate students in Exercise Science, Coaching, Athletic Training, and Athletics Sports programs that perform research in biomechanics and are interested in video processing for biomechanical analysis. The combination of student experiences using the existing Sport Analysis Center - Ariel Analysis Performance System motion analysis software and the proposed video streaming research provides students and scientists with a highly technical experiential opportunity that could distinguish them from individuals doing research in digitizing video for biomechanical analyses.

*Research Objectives and Questions*

The purpose of this research was to identify an optimal computer hardware model that helps in reducing the number of factors that contribute to the loss of video fields when 60 de-interlaced fields per second digital uncompressed video streaming is performed over IEEE 1394a connection to a Gateway 450ROG laptop with installed Ariel Performance Analysis System motion analysis software. Additionally, this study's goal was to provide statistical analysis of the number of video streaming sources and how various hardware components such as memory size or hard disk revolutions per minute (RPM) affect the problem with lost video fields. In other words the substantive research questions are stated as follows:

$Q_1$: Does hardware configuration such as memory (RAM) size or hard disk speed (RPM) contribute to the loss of video fields?

$Q_2$: Is the loss of video fields affected by the number of streamed video sources?

$Q_3$: Do variables such as RAM, hard disk RPM and number of video sources significantly predict the number of lost video fields?

## Motion Analysis in Biomechanics

The importance of determining if hardware components and a number of streaming video sources contribute to the loss of video fields comes from a wide variety of choices in Central Processing Units (CPUs), Intelligent Drive Electronics (IDE) configurations, memory sizes, number of DV camcorders and other available computer hardware. In the process of transmitting DV-quality video via IEEE 1394a protocol, there are many factors that contribute to the loss of video fields. When a DV-camcorder records an image, it is first stored on a DV videocassette or it could be transmitted to a PC real-time via IEEE 1394a connection. Once the computer receives the video stream, it is further processed. There is a possibility that the loss of video fields occurs before a video stream reaches a PC, however, the purpose of this research was to determine which computer hardware components, such as memory size, hard drive type or a number of streaming video sources contribute to the loss of video fields during a DV-quality video transmission over IEEE 1394a protocol. Different IEEE 1394a wires such as shielded, unshielded, magnetically shielded or unshielded may also contribute to the loss of video fields; however, this study does not address different IEEE 1394a cables – it is limited to the IEEE 1394a 10m cable made by Unibrain (part# 1638).

*Figure 1* presents a scenario where a video frame is lost during transmission. This emphasizes the importance of receiving full video stream without lost video fields. Part *a)* of *Figure 1* shows a dropped frame during a streamed video of the performing athlete. Part *b)* of *Figure 1* shows the resulting video stream after the frame #3 was dropped during transmission.

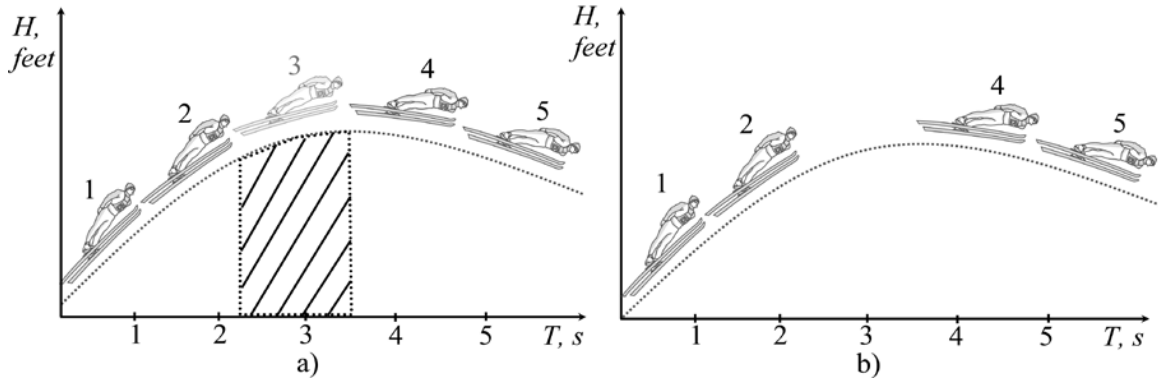

*Figure 1.* Video frame lost during transmission.

Once a situation depicted on the *Figure 1* occurs, quality digitizing of the video content is unlikely due to a missing video frame. Thus the video stream must be re-transmitted to complete the digitizing process. This described process is further complicated with a usage of three or more cameras capturing video content. A technique with three or more DV-camcorders is used when a 3-D analysis of video is desired. According to Allard, Blanchi and Aissaoui (1995), three-dimensional (3-D) biomechanical analyses "start with data capture by an imaging device". Object points are specified that allow for distinction between different video sources (Allard P., Strokes I., Blanchi J., 1995). Once a video is captured, then computer modeling of human movements is possible. The importance of computer simulation of human movement is stressed by Hamill and Whittlesey. They argue that computer models provide the following advantages (Robertson G., *et al.* 2004):

1. When dealing with computer models, the constraints associated with human subjects are eliminated: fatigue, strength and coordination, safety and ethics,

2. A computer model allows for experimenting with conditions that cannot be tested on human subjects,

3. Computer models can be adapted to search for optimal solutions.

Besides several advantages that computer models provide there are many limitations that remain, including (Robertson G., *et al.* 2004):

1. Numerical imperfections exist in the solution process,

2. Difficulty in modeling impacts,

3. Approximation of complex human structures limits computer models,

4. Computer models cannot be compared to enormous adaptability of the human system,

5. Assumptions are a must, since it is not possible to account for all human and environmental factors.

This research study employs the performance of the motion analysis software package called Ariel Performance Analysis Software (APAS). *Figure 2* shows the examples of how APAS could be used for biomechanical analysis of DV- quality video. In order to identify hardware components that contribute to the loss of video fields when 720x480 60 de-interlaced fields per second video is transmitted via IEEE 1394a connection to a laptop computer with installed Ariel Performance Analysis System motion analysis software, a comparative analysis of different computer hardware setups was needed.

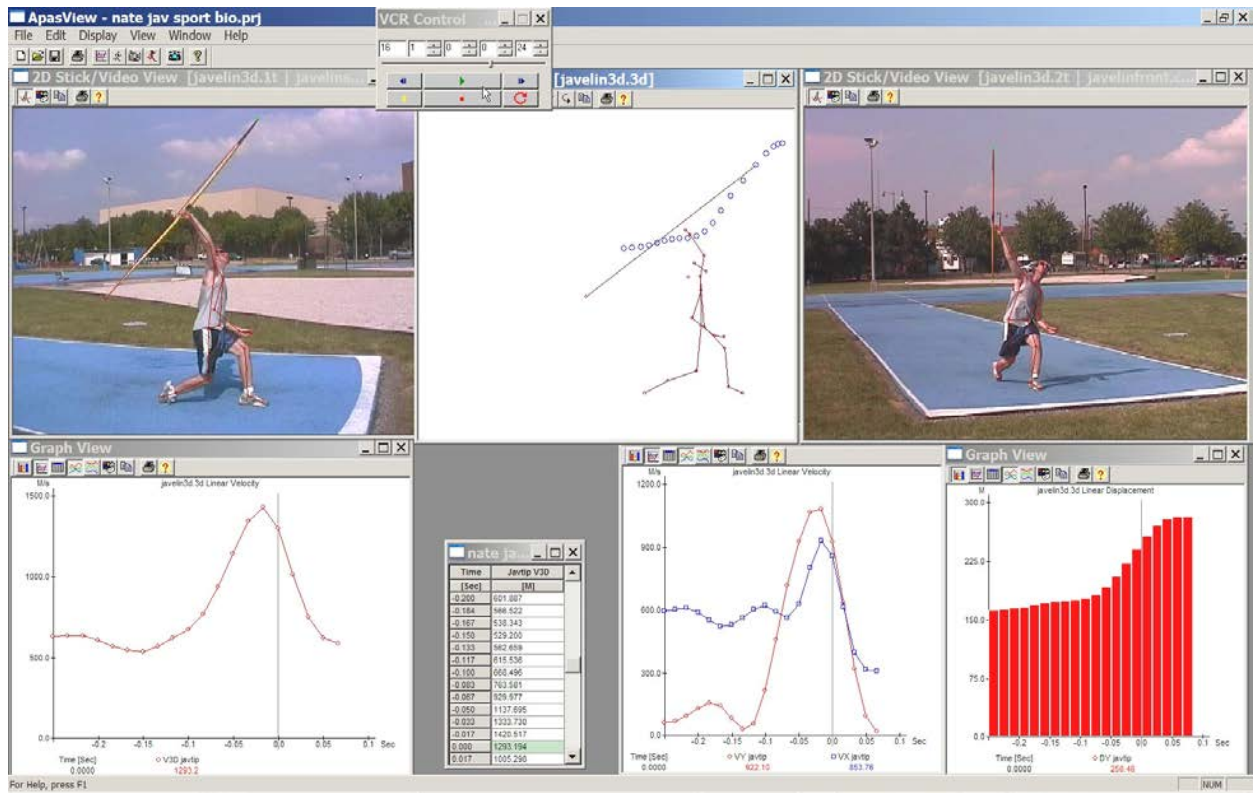

*Figure 2.* Sample of video based biomechanical analysis with multiple video views.

<div align="center">Research Design</div>

In order to evaluate the contribution of the hardware elements to the loss of video fields, several test scenarios were created and represented by Table 1. Each configuration test scenario A, B, C, D, E, or F represents the steps taken during the experiment.

Table 1

*Test configuration scenarios for Gateway 450ROG laptop*

|  | Configuration | | | | | |
|---|---|---|---|---|---|---|
|  | A | B | C | D | E | F |
| Memory Size, Mb | 1024 | 1024 | 1024 | 1024 | 1024 | 1024 |
|  | 512 | 512 | 512 | 512 | 512 | 512 |
| Hard Drive Speed, RPM | 7200 | 7200 | 7200 | 5400 | 5400 | 5400 |
| Number of Video Sources | 1 | 2 | 3 | 1 | 2 | 3 |

For example, in jumping events there are three attempts per heat with 10-12 athletes participating (30-36 trials). That creates thirty 10-second collections for each test configuration scenario A, B, C, D, E or F.

There were two FireWire controllers used in this study – one COMPAQ PCMCIA Cardbus controller and another one was onboard embedded into the tested Intel chipset of the Gateway 450ROG laptop. For every testing configuration with 1 or 2 video sources COMPAQ PC card was used; once there were three cameras connected to the Gateway PC, the third camera was plugged into the onboard IEEE 1394a controller. Each time a video transfer of video occurred, APAS indicated whether a loss of video fields took place. Every transfer session results were inputted into the population sample. *Figure 3* indicates a sample screen captured from APAS after a video transfer is complete.

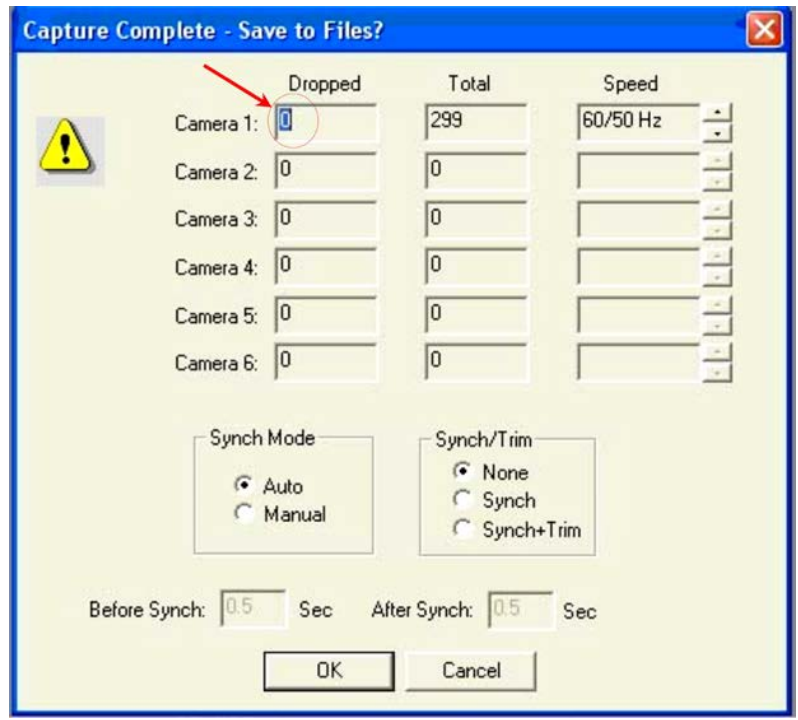

*Figure 3.* APAS indication of captured video fields.

The number of collections for each scenario is calculated in the following manner: a typical video transmission for 100 meter hurdles is expected to last about 12-15 seconds, so 10 seconds is used as a recording time for data collection. Therefore a factorial ANOVA design of 2 x 3 x 2 x 30 (RAM, RPM, VideoStreams, trials) with repeated measures on all factors will be used to analyze the number of dropped de-interlaced fields from the video streams (DV camcorders). *Figure 4* presents the statistical design used in this study which consists of 30 trials each lasting 10 seconds for two de-interlaced fields per second that will be collected for the twelve condition cells.

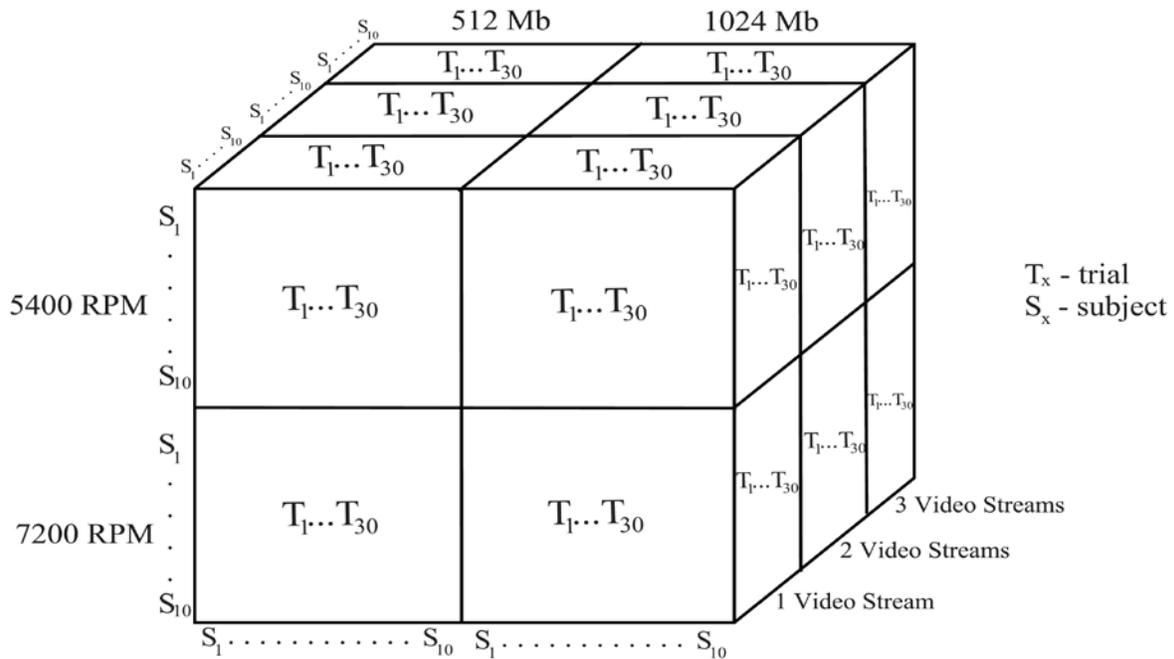

*Figure 4.* 2 x 3 x 2 x 30 factorial design.

*Instruments of Data Collection*

The following necessary tools and applications were utilized during the data collection process:

1. APAS - Ariel Performance Analysis System motion analysis software, version 2005. This software would allow analyzing video on the laptop computer transferred through the IEEE 1394a connection from the a DV camcorder(s) (video sources),

2. Acronis True Image Home version 9.0. This application's usage is for cloning an image of a hard drive,

3. Gateway 450ROG laptop with Windows XP Professional 32-bit version,

4. Hitachi hard drives HTS541060G9A00 (5400 RPM) and HTS726060M9AT00 (7200 RPM),

5. IEEE 1394a 10m cable made by Unibrain (part# 1638),

6. COMPAQ 2 Port Firewire PCMCIA Cardbus model no. CPQFWCB (serial # UMI4103952).

Findings

The purpose of this study was to identify hardware components that contribute to the loss of video fields when 720x480 60 de-interlaced fields per second video is transmitted via IEEE 1394a connection to a laptop computer with installed Ariel Performance Analysis System motion analysis software. There were several variables that were considered in this study – hard disk speed (RPM), memory (RAM) and the number of video sources (VideoStreams) connected to the tested system. Based on the results of this study, the amount of RAM did not have a significant main effect of the number of video fields transmitted.

Based on the results of this study, RPM variable showed a significant main effect on the number of collected video fields. Tested configurations with 7200 RPM hard drive showed lower performance in video frames collection. Five out of six tested configurations with RPM measure equal to 5400 did not result in any dropped fields. Only one tested PC configuration with 1024 Mb of RAM and 5400 RPM resulted in dropped fields. However, 2 out of 6 tested Gateway computer configurations with 7200 RPM and 512 Mb of RAM resulted in dropped fields. Furthermore, most resourceful configurations with 1024 Mb of RAM and 7200 RPM hard disk speed resulted in collections of more than expected 300 frames. That phenomenon occurred when more than 1 video camera transmitted video to the tested Gateway laptop.

Based on the results of this study, the number of video sources connected to the tested system had influenced the number of collected fields. The power coefficient for Video sources was calculated to be 1.00 which indicates an adequate sample size for the experiment. As for the power coefficients for the RAM and RPM, the resulted calculations showed a need for a larger sample size.

This study's research questions were answered in the following manner:

$Q_1$: Does hardware configuration such as memory (RAM) size or hard disk speed (RPM) contribute to the loss of video fields?

The hardware such as hard disk speed (RPM) contributes to the loss of video fields.

$Q_2$: Is the loss of video fields affected by the number of streamed video sources?

The loss of video fields is affected by the number of streamed video sources.

$Q_3$: Do variables RAM, hard disk RPM and number of video sources significantly predict the number of lost video fields?

Variables such as hard disk RPM and number of video sources significantly predict the number of lost video fields.

<div align="center">Recommendations</div>

Motherboards with chipset that are designed for mobile CPU chips allow for adjustable CPU clock speeds based on the immediate demands for processing power (Stepping Technology). The tested Gateway portable computer had an x86 Family 6 Model 13 Stepping 6 Genuine Intel ~1798 MHz processor installed that supports Stepping Technology. Perhaps, the CPU's Stepping Technology could explain why the most resourceful configurations of this study's research experiment with 1024 Mb of RAM and 7200 RPM hard disk speed resulted in collections of more than expected 300 frames. It is a possibility that the tested CPU switched to a faster clock speed when multiple video cameras were connected to the tested Gateway computer. This indicates a need to further investigate the effects of different CPU technologies' impacts on the number of collected video fields. For future research it is recommended to test the following CPU hardware configuration scenarios:

1. Test a CPU with Stepping Technology support vs. a regular, clock speed non-adjustable CPU,

2. Test a CPU with Stepping Technology support vs. a multi-core CPU According to Technology@Intel Magazine, "dual-core processor-based PCs are the next generation in PC computing performance and power Dual-core processor-based PCs are the next generation in PC computing performance and power" (Intel Corp. 2005),

3. Test the effectiveness of various levels of on-chip cache memory.

Another computer component that might affect the transmission of video to APAS software via IEEE 1394a connection is storage controller type. For example, Redundant Array of Independent Disks (RAID) is still widely utilized in industry; furthermore, home computers are starting to utilize this technology of RAID due to the decreasing pricing, better availability and a range of manufacturers that offer such technologies. In this study, Gateway computer utilized Intel(R) 82801FBM Ultra ATA Storage Controller – ATA-7 with 100Mb/s throughput. However, more resourceful systems with multiple hard drives that are configured in RAID could have a different main effect on the collected video fields. It is recommended for future research to establish whether faster and better performing hard drives and storage controllers affect the problem with lost video fields. The following are some recommended hardware testing configurations for computer hard drives:

1. Test SATA type I storage controller vs. SATA type II storage controller,

2. Test SATA type II storage controller vs. IDE ATA storage controller,

3. Test the impact of having different cache memory levels for hard drives – 2 Mb vs. 8 Mb, 8Mb vs. 16 Mb on the collection of video frames.

Another recommendation for future research is to test different FireWire chipsets that might contribute to the loss of video fields. Here are some suggestions:

1. Test FireWire 400 vs. FireWire 800 PC card types,

2. Test FireWire 400 vs. Dual Channel FireWire 400 PC card types.

In conclusion, it is hoped that completing this research study's objectives are found to be practical, resourceful and constructive. Future research studies should further identify an optimal computer hardware model that results in reducing the number of factors that contribute to the loss of video fields when 60 de-interlaced fields per second digital uncompressed video streaming is performed over IEEE 1394a connection.